\documentclass[review]{elsarticle}

\usepackage{lineno,hyperref}
\modulolinenumbers[5]

\usepackage{bbm}
\usepackage{graphicx}
\usepackage{amsmath}
\usepackage{amsfonts,amsbsy}
\usepackage{amssymb}

\newcommand{\be}{\begin{equation}}
\newcommand{\ee}{\end{equation}}

\newcommand{\st}{{\mathbf{s}_\perp}}

\newcommand{\bt}{{\mathbf{b}_\perp}}

\begin{document}
\title{Intrinsic Fluctuations of the Proton Saturation Momentum Scale in High Multiplicity p+p Collisions}
\author[bnl,ccnu]{Larry McLerran}
\author[bnl]{Prithwish Tribedy}

\address[bnl]{Physics Dept, Bdg. 510A, Brookhaven National Laboratory, Upton, NY-11973, USA}
\address[ccnu]{Physics Dept, China Central Normal University, Wuhan, China}

\begin{abstract}
High multiplicity events in p+p collisions are studied using the theory  of the Color Glass Condensate. We show that intrinsic fluctuations of the proton saturation momentum scale are needed in addition to the sub-nucleonic color charge fluctuations to explain the very high multiplicity tail of distributions in p+p collisions. The origin of such intrinsic fluctuations are presumably non-perturbative in nature. Classical Yang Mills simulations using the IP-Glasma model are performed to make quantitative estimations. We find that fluctuations as large as {$\cal O$}(1) of the average values of the saturation momentum scale can lead to rare high multiplicity events seen in p+p data at RHIC and LHC energies. Using the available data on multiplicity distributions we try to constrain the distribution of the proton saturation momentum scale and make predictions for the multiplicity distribution in 13 TeV p+p collisions.

\end{abstract}


\maketitle

\section{Introduction}
High multiplicity events in the collisions of small systems like p+p have recently become an important topic of interest. Such processes can provide important insights about the dynamics of multi-particle production arising from the saturated gluon fields inside the protons. However, the origin of such high multiplicity events from the first principles QCD approach is not yet fully understood. We demonstrate in this paper that the description of multiplicity fluctuations requires a framework of multi-particle production that includes several non-perturbative scales in the problem. 

Recent efforts to understand multpilicity fluctuations have  focused on applying the theory of the Color Glass Condensate (CGC)~\cite{McLerran:1994ni,McLerran:1994ka,Iancu:2003xm,Gelis:2010nm,Kovchegov:2012mbw}. This theory  successfully describes several features of the global data in hadronic, light-heavy, and heavy ion collisions. In the CGC framework, the emergence of a dynamical scale $Q_{S}$ makes the computation of multi-particle production feasible at very high energies. This is because the large values of $Q_{S}$ make the effective coupling small, thereby making weak coupling methods applicable. $Q_{S}$ being the only scale in the problem also naturally sets the scale of gluon field fluctuations that eventually give rise to the fluctuations in multiplicity. It turns out that using a single framework one can qualitatively describe both the bulk multiplicity and its fluctuations up to the level of uncertainties introduced by the knowledge of the saturation scales of the colliding systems. A quantitative description of the multiplicity fluctuations requires more sophisticated treatment of the non-perturbative dynamics of multi-particle production, modeling of the details of wave functions of the collision systems, and fluctuations in the geometry of initial stages of collisions~\cite{Tribedy:2010ab, Tribedy:2011aa, Dumitru:2011wq, Dumitru:2012yr}. 

The IP-Glasma model that combines the IP-Sat dipole model and non-linear dynamics of the Glasma gluon fields can provide an {\it ab inito} approach to such problems in the framework of the CGC~\cite{Schenke:2012wb,Schenke:2012fw}.

\section{Different Sources of Fluctuations in CGC : the IP-Glasma model}
The input to the IP-Glasma model is the distribution of the saturation scale in the co-ordinate space transverse to collision direction at a given energy, $Q_S^2(\st)$, inside a proton from the IP-Sat model~\cite{Kowalski:2003hm}. The transverse distribution of $Q_S^2(\st)$ is controlled by the thickness function of the proton, which in the IP-Sat model has a Gaussian form given by
\be
T_p(\st) = \frac{1}{2 \pi B_G} \exp \left(\frac{-\st^2}{2 B_G}\right), 
\label{eq_tp}
\ee 
where the parameter $B_G= 4$ GeV$^{-2}$ that determines the width of the gluon distribution is obtained from fits to HERA data~\cite{Kowalski:2003hm, Rezaeian:2012ji}. The transverse distribution of $Q_S^2(\st)$ inside a proton obtained this way is a smooth distribution. This distribution of $Q_S^2(\st)$ can be used to sample different configurations of the color charges inside the proton according to the McLerran-Venugopalan (MV) model~\cite{McLerran:1994ni,McLerran:1994ka} using a relation between $Q_S^2(\st)$ and the variance of the color-charge distribution $g^2\mu(\st)$ in the MV model~\cite{Lappi:2007ku}. This is how sub-nucleonic fluctuations are introduced in the IP-Glasma model. The distributions of the gluon fields inside two colliding protons are obtained by solving Classical Yang-Mills equations in 2+1 D lattice. Via the solutions of these equations, the 
 original fluctuations in the configuration of color charges get transmitted to the gluon fields of the two colliding protons. 

The second source of the fluctuations in the IP-Glasma model is of geometric origin. The gluon fields produced after the collision of two protons are obtained in terms of the gluon fields of the two colliding protons. The resultant field after the collision vanishes in the region where the fields of the two protons do not overlap. Therefore an additional source of stochasticity is introduced by the fluctuations in the overlap area of the two protons due to fluctuation in the impact parameter of the collision. The distribution of the fluctuation of the impact parameter in p+p collisions is not known form first principles. However, it can be well approximated according to an eikonal model for p+p collisions. In that approach, one calculates the overlap function of the two protons at a given impact parameter $b=|\bt|$, given by 
\be
T_{pp}(b) =\int {\rm d}^2\st  \, T_p^A(\st)\, T_p^B(\st-\bt). 
\label{eq_tpp} 
\ee
Then one defines the differential probability of a p+p collision with impact parameter $b$ as ~\cite{d'Enterria:2010hd} 
\be
\frac{dP}{d^2b}(b) = \frac{1-e^{-\sigma_{\rm gg}N_g^2 T_{pp}(b)}}{\int {\rm d}^2b \left(1-e^{-\sigma_{\rm gg}N_g^2 \, T_{pp}(b)}\right)},
\label{eq_dpdb}
\ee
where $N_g^2 \sigma_{\rm gg}$ is the total effective partonic cross section at a given energy of collisions. The value of $N_g^2 \sigma_{\rm gg}$ is adjusted in such a way that the denominator of Eq.~\ref{eq_dpdb} becomes the inelastic p+p collision cross section $\sigma_{_{\rm NN}}^{\rm inel}$ at that collision energy. To estimate the value of $\sigma_{_{\rm NN}}^{\rm inel}$ for different energies $\sqrt{s}$ (in GeV) one uses the parameterization obtained from fits to the global data of total ~\cite{Agashe:2014kda} and elastic~\cite{Antchev:2011vs} cross sections given by 
\be
\sigma_{_{\rm NN}}^{\rm inel} (\sqrt{s}/ 1{\rm \, GeV}) = 25.2 + 0.05 \ln(\sqrt{s}) + 0.56 \ln^2(\sqrt{s}) + \frac{45.2}{\sqrt{s}^{0.9}} + \frac{33.8}{\sqrt{s}^{1.1}} {\rm \,\,\,\,\,\, mb}.
\label{eq_siginel}
\ee
This particular ansatz for implementing event-by-event fluctuations of the impact parameter in p+p collisions was proposed in Ref~\cite{d'Enterria:2010hd} and adopted in the IP-Glasma model~\cite{Schenke:2013dpa}.

The combined effect of these two sources of fluctuations leads to event-by-event fluctuations in the transverse distribution of the gluon fields produced by the collisions. In the IP-Glasma model, these gluon fields are evolved in time $\tau$ according to the Classical Yang-Mills equations. The gluon multiplicity is calculated after an evolution time of $\tau \sim 1/Q_S$ when the fields becomes weakly interacting and a freely streaming system of gluons. The gluon multiplicity obtained in such a way fluctuates from event to event due to the combined effect of the sub-nucleonic and geometric fluctuations.

The Gaussian fluctuations of the color charges in the MV model give rise to a negative binomial distributions (NBDs) of gluon number fluctuations for a given overlap geometry of collisions. This was first shown in a perturbative computation of the Glasma flux-tube picture~\cite{Gelis:2009wh} and was found to be naturally incorporated in the non-perturbative framework of the IP-Glasma model~\cite{Schenke:2012fw}. Due to event-by-event fluctuation of the impact parameter, the resultant distribution of multiplicity is a convolution of multiple NBDs each weighted by the corresponding values of impact parameter given by Eq.(\ref{eq_dpdb}). 

In Ref.~\cite{Schenke:2013dpa} it was shown that these two sources of multiplicity fluctuations are not sufficient for explaining the width of the experimental multiplicity distribution of charged particles in p+p collisions. The resultant distribution from the IP-Glasma model was found to be too narrow as compared to the data. After introducing additional fluctuations of $Q_S(\st)$ according to a Gaussian distribution, the multiplicity distribution was found to be closer to data. As we discuss in the next section, the inclusion of such fluctuations has been pointed out in several previous treatments of the CGC and is the main topic of this paper.

\section{Additional Sources of Fluctuations} 
Incorporating event-by-event fluctuations of the saturation momentum of the proton at a given energy of collisions goes beyond the conventional framework of CGC. 
The importance of such fluctuations to explain the pseudo-rapidity dependence of multiplicity in p+A collisions has been highlighted in a very recent work~\cite{McLerran:2015lta}. It has been clearly demonstrated that large fluctuations in the saturation scale of the proton are necessary to get the right slope of the pseudo-rapidity dependence of multiplicity for different centralities.
 The intrinsic fluctuation of the saturation momentum in the CGC framework has also been investigated in Ref.~\cite{Dusling:2013qoz, Dusling:2012iga}. A large increase of the intrinsic saturation scale of protons in high multiplicity events was shown to be essential for a quantitative description of long range two-particle ridge correlations in the CGC framework~\cite{Dusling:2013qoz,Dusling:2012iga}. In was shown that in p+p collisions a fluctuation of proton saturation scale $Q_S^2$ by about 5-6 times its min-bias value is required to model rare high multiplicity events in which the ridge like correlation was measured in the experiment. We will revisit this point in a later section of the manuscript. 
In this work we try to demonstrate the importance of such fluctuations for the description of the n-particle multiplicity distribution in p+p collisions. 

The idea of intrinsic fluctuations of the saturation momentum goes back to Ref.~\cite{ Iancu:2004es, Iancu:2004iy}, which extended a correspondence between high-energy QCD and the reaction-diffusion problem that was originally identified in Ref.~\cite{Munier:2003sj, Munier:2003vc} at the level of mean field approximation. The origin of such fluctuations is better understood within the color dipole picture ~\cite{Mueller:1993rr, Mueller:1994gb, Iancu2004494}. In the large $N_c$ limit the gluon distribution inside a target proton at a given rapidity $Y$ can be thought of as a distribution of dipoles. This picture is however valid only in the dilute regime of the target, in the saturation regime the dipole picture breaks down. In this picture one can therefore define the scattering amplitude of the target $T($$r)$ for a dipole of size $r$. The dilute regime of the target corresponds to small values of $r\rightarrow0$ where the distribution $T($$r)$ has a long tail approaching zero. In the saturation regime (large $r$) where the dipole picture breaks down $T($$r)$ approaches the unitarity limit. The saturation scale of the target at that rapidity $Q_S(Y)$ is defined as the inverse of the dipole size $r$ for which $T($$r=1/Q_S(Y)) = T_S$, where $T_S$ is of the order of unity~\footnote{The exact value for the choice of $T_S$ is not important, it can only introduce a logarithmic uncertainty, it is often taken to be $e^{-1/2}$ }. Therefore the value of $Q_S$ is determined by the tail of the distribution $T($$r)$ which corresponds to the dilute regime of the target. In the dipole picture the natural variable characterizing the size of a dipole is expressed as $\ln(1/r^2)$, therefore the tail of $T($$r)$ is actually related to the logarithm of saturation scale $\ln(Q_S^2)$. 

With the evolution in rapidity $Y$, individual dipoles inside the target start to split into new dipoles. This nature of the splitting of individual dipoles is a purely stochastic process. The result of such splitting would change the distribution of the scattering amplitude for a starting value of rapidity $Y_0$ given by $T($$r,Y_0)$ to an evolved distribution at rapidity $Y$ given by $T($$r,Y)$. 

Due to the stochastic nature of the dipole splitting, the distribution $T($$r,Y)$ will have a different shape if the evolution is repeated for a second time starting with the same initial distribution $T($$r,Y_0)$. In the conventional framework of CGC, the evolution of the dipole scattering amplitude is implemented by BK or JIMWLK renormalization equations~\cite{Balitsky:1995ub, PhysRevD.61.074018, Jalilian-Marian:1997xn, Iancu:2000hn, Ferreiro:2001qy} . However the effect of this stochastic nature of the dipole splitting is not incorporated in either of the two evolution equations. These two equations only incorporate the evolution of the average distribution of the dipole scattering amplitude $\langle T(\,r\,) \rangle_Y$. This was initially pointed out by the authors of Ref~\cite{Iancu:2004iy}. The stochastic nature of the dipole splitting therefore would lead to multiple distribution of $T($$r,Y)$ for a given initial distribution of the scattering amplitude $T($$r,Y_0)$. Since the tail of such a distribution determines the effective value of the saturation momenta, this would lead to a distribution of $\ln(Q^2_S(Y))$ after an evolution of rapidity $Y$ around a mean value of $\ln(\langle Q^2_S(Y) \rangle)$. In reference ~\cite{Iancu:2004es} it was shown that the variance $\sigma^2$ of such distribution would depend on steps of evolution in rapidity and the strength of coupling. The evolution of $\sigma^2$ was found to be linear in rapidity. However the absolute value of $\sigma^2$ was argued to have a non-perturbative origin. The effect of non-zero $\sigma^2$ was conjectured to be incorporated in a Gaussian distribution describing the fluctuation of $\ln(Q^2_S)$~\cite{Iancu:2004es}. 
The explicit expressions of the cumulants of such fluctuations in the context of a stochastic reaction-diffusion model has been derived in Ref~\cite{PhysRevE.73.056126}. The full derivation of the distribution of saturation scale has been done in Ref~\cite{Marquet:2006xm}. It was shown that the leading behavior of such distribution is Gaussian in $\ln(Q^2_S(Y))$ and the correction to the Gaussian fluctuations is negligible at large $Y$.

In this work we introduce such event-by-event Gaussian fluctuations of the saturation momentum of the two colliding protons in the IP-Glasma model and study the effect on multiplicity fluctuations in p+p collisions at RHIC and LHC energies.

\section{Results}

We employ the IP-Glasma model to estimate the event by event distribution of multiplicity. The parameterization for the saturation scale in the IP-Sat model is obtained from the recent fits to the HERA data performed in Ref~\cite{Rezaeian:2012ji}. In this work we perform the simulation for p+p collisions at five different energies $\sqrt{s}=$ 0.2, 0.9, 2.36, 7 and 13 TeV. The details of the lattice implementations of the IP-Glasma model can be found in Ref.~\cite{Schenke:2013dpa}. The only difference from Ref.~\cite{Schenke:2013dpa} is that in this work we use slightly finer lattices with a transverse size of L=8 fm with N=400 cells in each direction. All other parameters related to the lattice implementations of IP-Glasma model are kept same to what have been used in Ref~\cite{Schenke:2013dpa} for the study of p+p collisions. 

In every event we allow the value of $Q_S^2(\st)$ at every point in the transverse plane to fluctuate according to the distribution~\cite{Iancu:2004es} given as
\be
P(\ln(Q_S^2/\langle Q_S^2 \rangle)) =\frac{1}{\sqrt{2\pi}\sigma}\exp\left(  -\frac{\ln^{2}(Q_{S}^2(\st)/\langle Q_{S}^{2}(\st) \rangle)}{2\sigma^{2}}\right).%
\label{eq_qsdist}
\ee
\begin{figure}[t]
\includegraphics[width=0.6\textwidth]{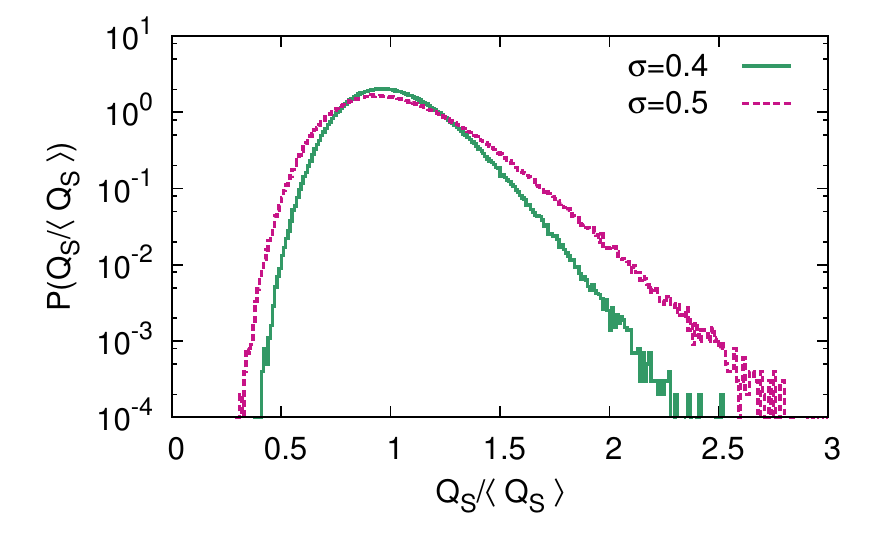}
\caption{Distribution of saturation scale of the proton as described by Eq.(\ref{eq_qsdist}) shown for two different values of the width $\sigma$. The tails of such distributions correspond to rare  configurations of the proton with large values of saturation momentum.}
\label{qs_fluc}
\end{figure}
This particular choice of the distribution gives rise to a skewed distribution of $Q_S/\langle Q_S \rangle$ around unity. The distribution of $Q_S(\st)$ for different values of $\sigma$ is shown in Fig.\ref{qs_fluc}. For simplicity, we assume $\sigma$ to be independent of the transverse distance $\st$ inside the proton~\footnote{In principle $\sigma$ should depend on the gluon density that varies with $\st$}. However as described in Ref~\cite{Iancu:2004es}, we assume the following rapidity dependence of $\sigma$ 
\be
\sigma^2 (Y) = \sigma_0^2 (Y_0)  + \sigma_1^2 (Y-Y_0), 
\label{eq_sigma}
\ee 
where $Y$ and $Y_0$ denotes the final and initial rapidity defined as $Y=\ln(2 \sqrt{s}/1{\rm GeV})$. Here we keep terms up to the first order in the expansion of rapidity. The contribution to $\sigma$ that is purely due to quantum evolution in rapidity is controlled by $\sigma_1$, whereas the origin of $\sigma_0$ is generally assumed to be non-perturbative. In our context both $\sigma_1$ and $\sigma_0$ are treated as free parameters and our goal is to constrain them using the experimental multiplicity distributions. 

A given distribution of $Q_S$ sets the color charge density $g^2\mu$ in the IP-Glasma model. 
The ratio of saturation scale to the color charge density, $Q_S(\st)/g^2\mu(\st)$ is treated as a parameter which is to be extracted from data. A value of $Q_S(\st)/g^2\mu(\st)=0.65$ was previously found to provide the good fit to the multiplicity distribution of different collision systems. However a slight discrepancy in terms of explaining the long tails of the probability distributions of multiplicity in case of p+p collisions was found in Ref~\cite{Schenke:2013dpa}. We therefore study the dependence of our results on the variation of this parameter in the following section and see if this parameter can be further constrained by the experimental distribution of multiplicity in p+p collisions. 

With this initialization we compute the gluon multiplicity per unit rapidity at rapidity $y=0$ after an evolution time of $\tau=0.4$ fm/c. We do not use any prescription of fragmentation and assume the gluon numbers to be proportional to the number of charged particles. In this work we express our results in terms of scaled multiplicity $N/\langle N \rangle$ to minimize the uncertainties related to the absolute normalization of the multiplicity. 

\begin{figure}[htb]

\hspace{-13pt}\includegraphics[width=0.53\textwidth]{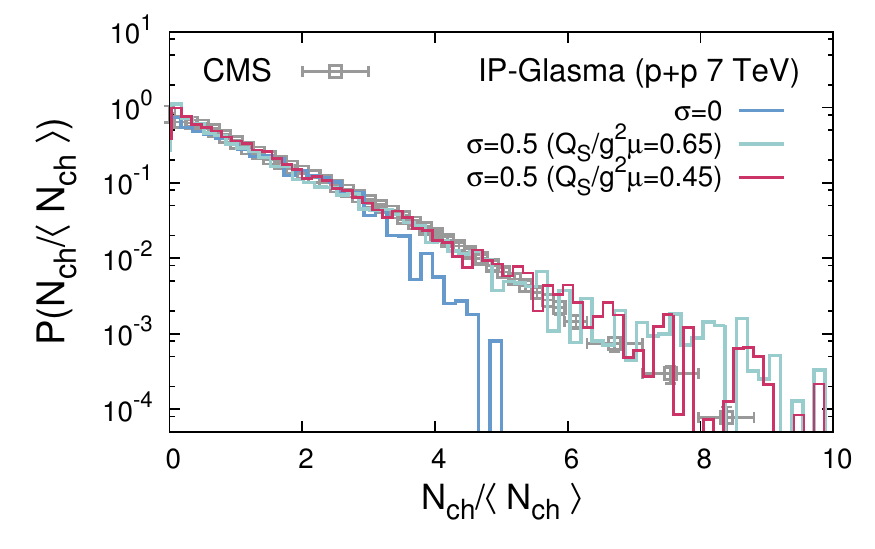}
\hspace{-5pt}
\includegraphics[width=0.53\textwidth]{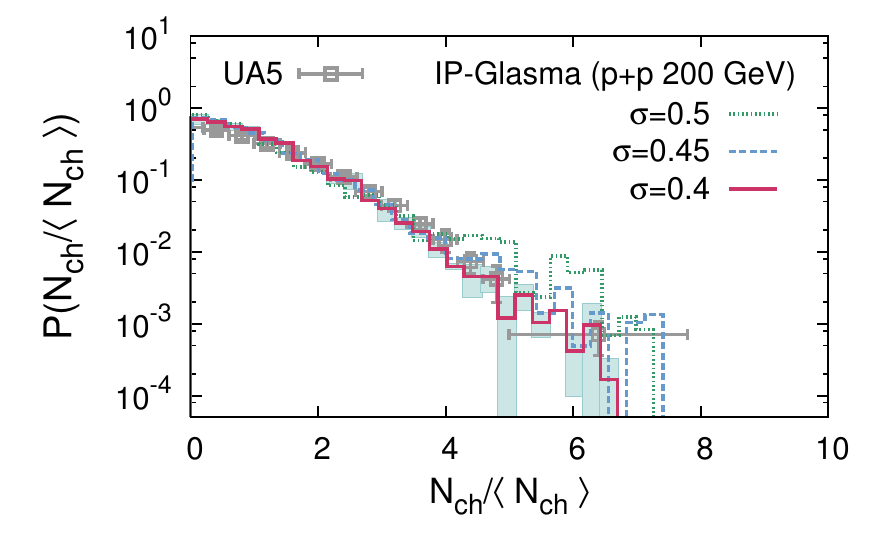}
\caption{Left: Probability distribution of scaled multiplicity in p+p collisions for 7 TeV. Experimental data points are from Ref.~\cite{Khachatryan:2010nk}. The two curves correspond to two different value of $\sigma$ parameters. Right: The same plot for p+p collisions at 200 GeV. The green band denotes the uncertainties due to the variation of  $Q_S/g^2\mu$ from 0.65 to 0.45. The experimental data points are for the pseudo-rapidity range of $|\eta|<0.5$ and obtained from Ref.~\cite{Ansorge:1988kn,Aamodt:2010ft}.}
\label{mult_7tev}
\end{figure}
The minimum bias distribution of scaled multiplicity for p+p collisions at 7 TeV is shown in Fig.\ref{mult_7tev} (left). The multiplicity distribution obtained from the IP-Glasma model for two different values of $\sigma$ are shown by the histograms. It is very much evident from Fig.\ref{mult_7tev} that in the absence of $Q_S$-fluctuations the estimated distribution (corresponding to $\sigma=0$) is much narrower compared to the data. This result is consistent to the previous findings of Ref~\cite{Schenke:2013dpa}. The inclusion of $Q_S$-fluctuations with $\sigma=0.5$ leads to a much better description of the data up to the values of scaled multiplicity $N_{\rm ch}/\langle N_{\rm ch} \rangle \sim$ 6. For $N_{\rm ch}/\langle N_{\rm ch} \rangle >$ 6 we find that the using a ratio of $Q_S/g^2\mu=0.45$ instead of 0.65 gives slightly better agreement to the data. A possible origin of such dependence on $Q_S/g^2\mu$ can be traced back to Ref~\cite{Schenke:2012fw} where the non-perturbative corrections to the Glasma flux-tube picture were found to be dominant in the regime of small $g^2\mu$ (or large $Q_S/g^2\mu$). For large values of $g^2\mu$ where the Glasma flux-tube picture is recovered the results will be insensitive to the choice of $Q_S/g^2\mu$. The fact that the IP-Glasma multiplicity distribution is sensitive to the choice of $Q_S/g^2\mu$ indicates that in p+p collisions we are still in a regime where non-perturbative corrections are important and the experimental data need to constrain the ratio $Q_S/g^2\mu$.
\begin{figure}[htb]
\hspace{-13pt}\includegraphics[width=0.53\textwidth]{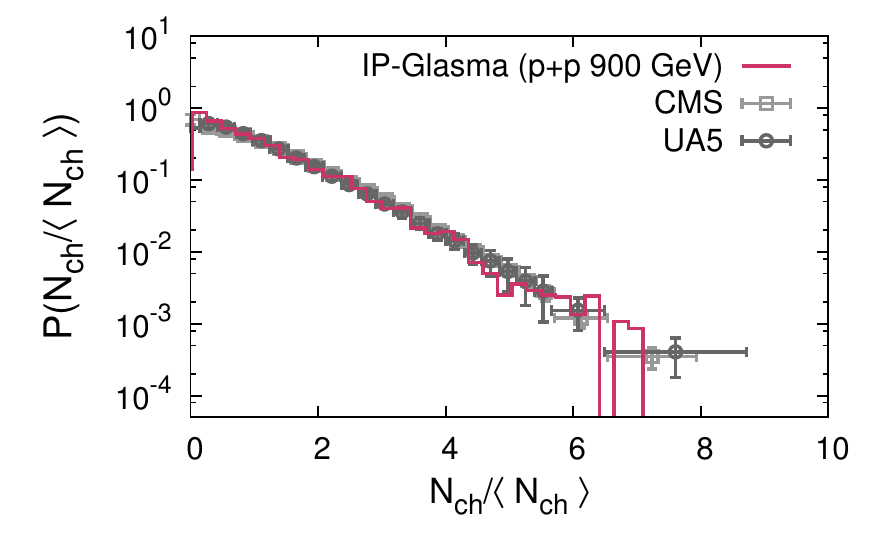}
\hspace{-5pt}\includegraphics[width=0.53\textwidth]{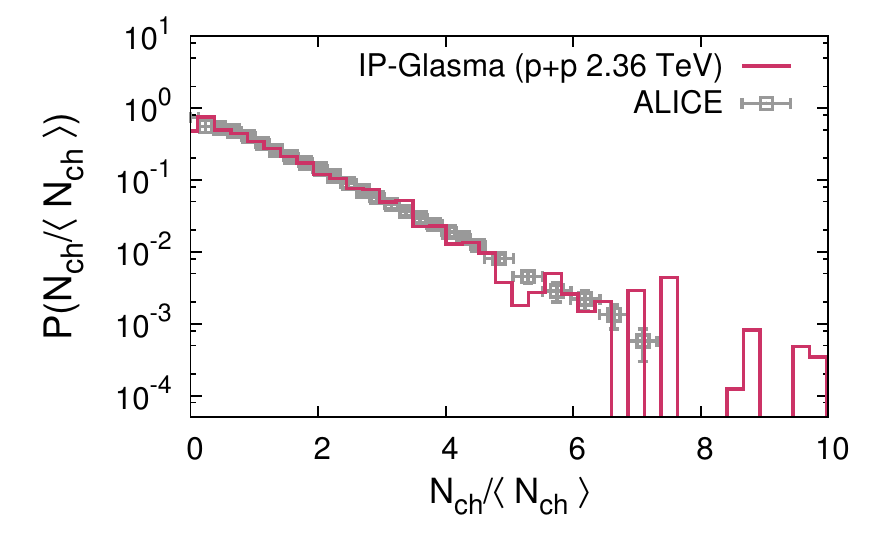}
\label{mult_900gev}
\caption{Left: Probability distribution of scaled multiplicity in p+p collisions for 900 GeV. Right: The same plot for p+p collisions at 2.36 TeV. The experimental data points are for the pseudo-rapidity range of $|\eta|<0.5$ and obtained from Ref.~\cite{Khachatryan:2010nk,Aamodt:2010ft}.}
\end{figure}

In Fig.\ref{mult_7tev} (right) we show the same distribution for 200 GeV. In this case we use three different values of $\sigma=$ 0.4, 0.45 and 0.5. We do not observe significant differences in the shape of the distributions for $\sigma=$ 0.4 and 0.45, the distribution corresponding to 0.5 slightly overestimates the data. For $\sigma=$ 0.4, we show the dependence of the distribution on the variation of the ratio $Q_S/g^2\mu$ in the range of 0.45 to 0.65 by a green band. Below $N_{\rm ch}/\langle N_{\rm ch} \rangle \sim$ 6  we see that the variation in the value of the parameter $\sigma$ that provide good fit to both the data (at 200 GeV and 7 TeV) must be less than 0.1. This would restrict the parameter $\sigma_1$ in Eq.(\ref{eq_sigma}) to be $\sigma_1^2 <0.025$ corresponding to the variation of rapidity $Y=$9.55 ($\sqrt{s}=7$ TeV) to $Y=$5.99 ($\sqrt{s}=200$ GeV). This negligible variation of $\sigma_1^2$ is consistent to the expectations of Ref.~\cite{Iancu:2004es}, indicating that the contribution to the $\sigma$ from evolution in rapidity (at small rapidity) is negligible as  compared to the dominant non-perturbative contribution $\sigma_0$. A maximum variation of $\sigma$ from 200 GeV to 7 TeV to be 0.1 as suggested by Fig.\ref{mult_7tev} leads to a dependence like $\sigma^2(Y) = 0.16 + 0.025 (Y-5.99)$. This would correspond to $\sigma \sim 0.45$ at $\sqrt{s}$=900 GeV and $\sigma \sim 0.47$ at $\sqrt{s}$=2.36 TeV. The corresponding distributions obtained for these two values of $\sigma$ are shown in Fig.\ref{mult_900gev}. Good agreement with data again confirms negligible dependence of $\sigma$ with energy. We would however like to emphasize the fact that since IP-Glasma simulations are numerically intensive we do not perform fine tuning of the $\sigma$ parameters. Based on the results shown in Fig.2 and Fig.3 we would like to restrict our conclusion to the fact that a Gaussian fluctuation with $\sigma \sim 0.5 \pm 0.1$ with weak energy dependence provides a good description of the multiplicity distribution in p+p collisions at RHIC and LHC energies. 
These results indicate that for rare events, that are one out of ten thousand, the proton will fluctuate into a configuration which correspond to a saturation scale $Q_S^2$ that is more than five times the magnitude of the average saturation scale. It is interesting to note that a similar order of fluctuation of the proton saturation scale was previously considered in Ref~\cite{Dusling:2013qoz, Dusling:2012iga} to model the high multiplicity events in which the signals of ridge correlations are observed in the experiments. 
%
%
%


%
\begin{figure}[htb]
\includegraphics[width=0.53\textwidth]{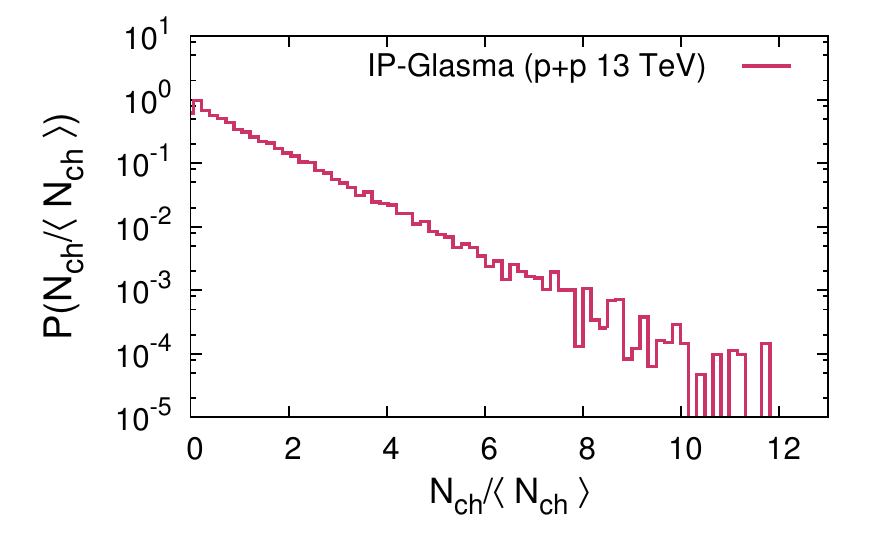}
\label{mult_13tev}
\caption{A prediction for the probability distribution of scaled multiplicity in p+p collisions for 13 TeV.}
\end{figure}

Finally we make a prediction for the recent 13 TeV p+p collisions at LHC using our framework. According to the parameterization of Eq.\ref{eq_sigma}, we expect the value of $\sigma\sim 0.51$ at 13 TeV. The inelastic cross section for 13 TeV is obtained from Eq.\ref{eq_siginel} to be $\sigma_{_{\rm NN}}^{\rm inel}(\sqrt{s}=13 {\rm \, TeV}) \sim 76$ mb. The estimated probability distribution for the scaled multiplicity is shown in Fig.\ref{mult_13tev}. In this case we have used the value of $g^2\mu/Q_S=0.45$. The experimental data of the corrected multiplicity distribution at 13 TeV will further constrain the parameter $\sigma$ and its energy dependence.

\section{Summary}
In this work have explored the origin of high multiplicity events in p+p collisions using the framework of CGC. The dominant source of multiplicity fluctuation in the framework of CGC is known to be the color charge fluctuations incorporated in the MV model, which with fluctuations of collision geometry can provide a qualitative explanation of the origin of the high multiplicity events. However, state of the art simulations of CGC using the IP-Glasma model demonstrate that additional sources of fluctuations are required to explain the origin of rare events which give rise to the tails of the multiplicity distributions in p+p collisions. In this work we demonstrate that including an additional source of fluctuations  in the intrinsic saturation momenta of a proton in p+p collisions, one can very well describe the multiplicity distributions over a wide range of energy. In the color dipole picture of high energy scatterings in QCD, such fluctuations are a consequence of stochastic splitting of dipoles that are not accounted for in the conventional frameworks of CGC. In this work we try to incorporate such process by fluctuating the intrinsic saturation momentum of two colliding protons in the IP-Glasma model according to a Gaussian distribution. The width of the Gaussian distribution $\sigma$ is an unknown parameter in this approach. We find that the multiplicity distributions over a wide range of energy form 200 GeV to 7 TeV is explained with a Gaussian of width $\sigma\sim0.5$. Two important conclusions that can be made from such observations are: (i) the fluctuations of ${\cal O}(1)$ in the saturation momentum are essential to explain the origin of high multiplicity events in the framework of CGC and (ii) the energy dependence of such fluctuations as constrained by the data appears to be negligible; that is, the fluctuations are dominantly of non-perturbative origin.

\section*{Acknowledgement} 
We would like to thank M. Praszalowicz for important discussions. We thank E. Iancu, C. Marquet, B. Schenke, S. Schlichting and R. Venugopalan for their important comments on the manuscript. The authors are supported under Department of Energy contract number Contract No. DE-SC0012704.

\bibliographystyle{elsarticle-num}
\section*{References}

\bibliography{qsfluctuation}

\end{document}